%% file: main.tex
\documentclass[nohyperref]{article}

\usepackage[accepted]{icml2022}

\input{packages}
\input{macros}

\theoremstyle{plain}
\newtheorem{theorem}{Theorem}[section]

\theoremstyle{definition}
\newtheorem{definition}[theorem]{Definition}

\newtheorem{property}[theorem]{Property}
\theoremstyle{remark}

\usepackage[textsize=tiny]{todonotes}

\begin{document}

\icmltitlerunning{Exploiting and Defending Against the Approximate Linearity of Apple's \nh}

\twocolumn[
\icmltitle{Exploiting and Defending Against the Approximate Linearity of \\ Apple's \nh}

\icmlsetsymbol{equal}{*}

\begin{icmlauthorlist}
\icmlauthor{Jagdeep S Bhatia}{mit,equal}
\icmlauthor{Kevin Meng}{mit,equal}
\end{icmlauthorlist}

\icmlaffiliation{mit}{Massachusetts Institute of Technology}

\icmlcorrespondingauthor{Jagdeep Singh Bhatia}{jagdeep@mit.edu}
\icmlcorrespondingauthor{Kevin Meng}{mengk@mit.edu}

\icmlkeywords{Perceptual hashing, cryptography, privacy, security, hashing, convolutional neural networks, genetic algorithms}

\vskip 0.3in
]



\printAffiliationsAndNotice{\icmlEqualContribution} 

\begin{abstract}
Perceptual hashes map images with identical semantic content to the same $n$-bit hash value, while mapping semantically-different images to different hashes.
These algorithms carry important applications in cybersecurity such as copyright infringement detection, content fingerprinting, and surveillance.
Apple's \nh is one such system that aims to detect the presence of illegal content on users' devices without compromising consumer privacy. We make the surprising discovery that \nh is \textit{approximately linear}, which inspires the development of novel black-box attacks that can (i) evade detection of ``illegal'' images, (ii) generate near-collisions, and (iii) leak information about hashed images, all without access to model parameters. These vulnerabilities pose serious threats to \nh's security goals; to address them, we propose a simple fix using classical cryptographic standards.
\end{abstract}

\input{report/intro}
\input{report/prelims}
\input{report/attacks}

\input{report/system-improvements}
\input{report/conclusion}




\bibliography{biblio}
\bibliographystyle{icml2022}



\end{document}

%% file: packages.tex
\usepackage{amsmath}
\usepackage{amssymb}
\usepackage{mathtools}
\usepackage{amsthm}

\usepackage{graphics}
\usepackage{amsfonts}
\usepackage{xspace}
\usepackage{latexsym}
\usepackage{enumitem}
\usepackage{graphicx}

\usepackage{microtype}
\usepackage{subfigure}
\usepackage{booktabs}
\usepackage{hyperref}

\hypersetup{
    colorlinks,
    linkcolor={red!50!black},
    citecolor={blue!50!black},
    urlcolor={blue!80!black}
}

%% file: macros.tex
\newcommand{\nh}{\textsc{NeuralHash}\xspace}
\newcommand{\expv}[1]{\mathbb{E}\left[ #1 \right]}
\newcommand{\expvsub}[2]{\mathbb{E}_{#2}\left[ #1 \right]}

\newcommand{\negl}{\mathrm{negl}}
\newcommand{\simfn}{\mathrm{sim}}
\newcommand{\ssim}{\mathrm{SSIM}}
\newcommand{\nha}[1]{\mathcal{N}\left(#1\right)}
\newcommand{\hashspace}{\mathcal{H}}

\definecolor{focus}{rgb}{0.77, 0.12, 0.23}

\let\[\equation
\let\]\endequation

%% file: report/intro.tex
\section{Introduction}

In 2021, Apple unveiled \nh, an algorithm aimed at detecting Child Sexual Abuse Material (CSAM) in images that are uploaded to iCloud \cite{apple-nh}.

\nh is a perceptual hash that, in general terms, aims to map images containing the same semantic information to the same hash, while mapping images with differing semantic content to different, random-looking hashes.
If executed correctly, such algorithms could potentially enable organizations to prevent illegal activity (e.g. copyright infringement, storing illicit audiovisual content) \textit{without compromising user privacy}. Thus, \nh is of great interest to the cybersecurity community.


Unfortunately, \nh has been controversial since its inception, due to concerns about efficacy and user privacy \cite{abelson2021bugs}. Recently, these concerns have been made concrete, with studies identifying problems with \nh's collision resistance, privacy guarantees, and false positive/negative rates under gradient-based adversarial attacks \cite{breaking-nh-struppek, nh-repro-github, nh-collisions-github, nh-collisions-blog}.


\nh relies on deep learning to hash images, in part because semantic image matching is difficult with rule-based or classical learning algorithms \cite{fuzzy-transforms, near-duplicate-im-detect}. Despite their advantages, however, neural networks are known to be sensitive to edge cases, susceptible to adversarial attacks, and prone to unexplainable behavior \cite{goodfellow2014explaining}.

White-box adversarial attacks on neural networks are well studied in literature; as a result, numerous gradient-based attacks have been proposed against \nh \cite{anish-github, breaking-nh-struppek}. We ask whether the same attacks can be achieved \textit{without} access to model weights, as the ability to construct such attacks in a gradient-free setting would imply a deeper understanding of \nh's inner workings. In particular, it is generally accepted that gradient-based attacks are an unavoidable phenomenon in deep neural networks \cite{ilyas-adv-examples-are-features}; the presence of strong \textit{black-box} attacks may suggest deeper structural flaws with \nh.

In this paper, we make the striking discovery that \nh is approximately linear in its inputs (Section \ref{sec:nh-linearity}), which not only violates \nh's privacy guarantees (Section \ref{subsec:lsh-def}), but also provides a framework under which we develop several strong black-box attacks (Section \ref{sec:attacks}). In Section \ref{sec:improvements}, we suggest a defense mechanism against the proposed attacks and demonstrate its effectiveness. To the best of our knowledge, we are the first to study the approximately linear nature of \nh's hashing mechanism.

\vspace{-5pt}


%% file: report/prelims.tex
\begin{figure*}[t]
  \centering
  \includegraphics[keepaspectratio, width=0.75\textwidth]{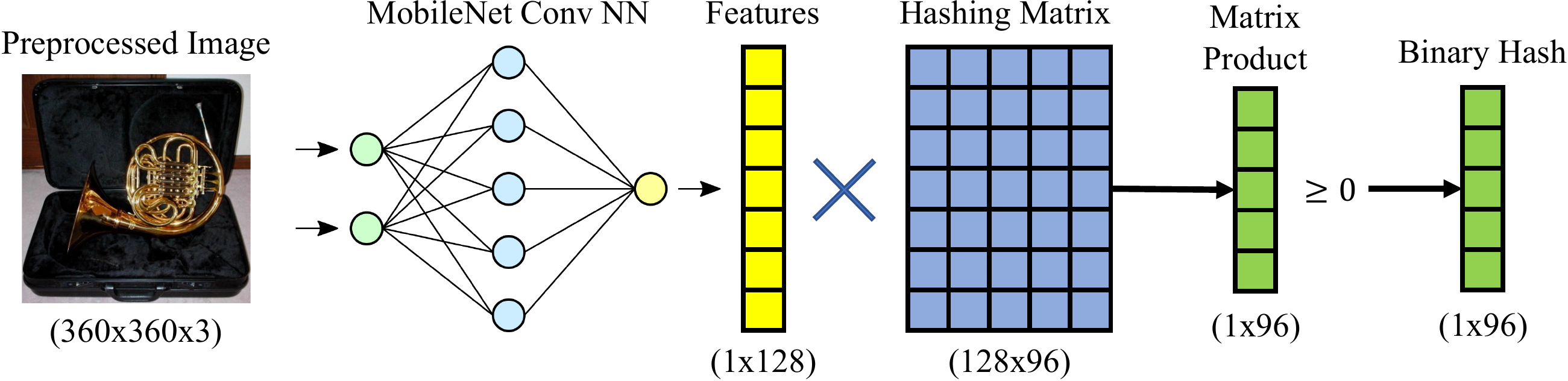}
  \caption{\textbf{\nh Pipeline}. The pre-processed image is first embedded with a contrastively-trained MobileNet \cite{howard2017mobilenets} convolutional neural network. The extracted features are then transformed with a randomized hashing matrix, whose outputs are passed through a step function to generate the final binary hash  \cite{breaking-nh-struppek, apple-nh}.}
  \label{fig:NH-Diag}
\end{figure*}

\section{Preliminaries} \label{sec:prelim}

\subsection{\nh Security and Privacy Goals} \label{subsec:lsh-def}

\begin{definition}
    Semantic Similarity: Two images $x_1, x_2 \in X$ are semantically similar if, given some metric function $d: X\times X \rightarrow \mathbb{R}$ and some threshold $\epsilon$, we have that $d(x_1, x_2) \leq \epsilon$. $d$ can be implemented in various ways, depending on the evaluation context.
\end{definition}
The primary difference between \nh and classical cryptographic hashing is that \nh aims to remain invariant under transformations that preserve semantic similarity. This gives rise to the first desired property of \nh:

\begin{property} \label{prop:neighborhood}
Neighborhood Consistency: Let $x_1, x_2 \in X$ be any semantically similar images and $\nha{x}: X \rightarrow \hashspace$ be the \nh perceptual hashing algorithm. It should be the case that $\nha{x_1} = \nha{x_2}$. Note that, in \nh, $\hashspace = \{0, 1\}^{96}$.
\end{property}

In analyzing the behavior of \nh, we adopt a variant of the Random Oracle Model that accounts for semantic similarity. In our setting, a Perceptual Random Oracle (PRO) should return a random hash for any previously-unseen query $x$, and the same hash value as a previous input $x^\prime$ if $x$ and $x^\prime$ are semantically similar:
\begin{equation}
    \nha{x} = \begin{cases}
        \nha{x^\prime} & \text{if $\exists x^\prime. \; d(x, x^\prime) \leq \epsilon$} \\
        h \sim \mathcal{H} & \text{otherwise.}
    \end{cases}
\end{equation}
Because we expect \nh to behave as an idealized PRO, it should satisfy the following properties:\footnote{Assume that adversaries are \textit{computationally bounded} and have unrestricted access to the hashing oracle $\nha{\cdot}$.}


\begin{property} \label{prop:uniform-hashing}
    Uniform Hashing: Let $x_1, x_2 \in X$ be any images that are \textit{not} semantically similar. It should be the case that (i) $\nha{x_1} \ne \nha{x_2}$ and (ii) the distributions over $\nha{x_1}$ and $\nha{x_2}$ look indistinguishable to a computationally bounded adversary. $\nha{x_1}$ and $\nha{x_2}$ should appear to be random elements of the hash space, and therefore,
    \begin{equation} \label{eq:uniform-hashing}
        \expvsub{\mathrm{sim}(\nha{x_1}, \nha{x_2})}{x_1, x_2 \in X, d(x_1, x_2) > \epsilon} \leq 0.5 + \negl,
    \end{equation}
    where, treating hashes as binary vectors,
    \begin{equation} \label{eq:hamming-sim}
        \mathrm{sim}(h_1, h_2) = \frac{h_1 \cdot h_2}{|h_1||h_2|}.
    \end{equation}
    Note that $\mathrm{sim}(h_1, h_2)$ is equal to one minus the normalized Hamming distance; call it \textit{Hamming similarity}.
\end{property}


\begin{property} \label{prop:weak-non-invert}
    Weak Non-Invertibility: Given target hash $h^* \in \hashspace$, it should be impossible for an adversary to generate any image $x$ such that $\nha{x} = h^*$.
\end{property}
\begin{property} \label{prop:strong-non-invert}
    Strong Non-Invertibility:
    Given any hash $h^* \in \hashspace$ and $n$ images $x_1, ..., x_n$ sampled uniformly at random from $X$, let $s$ be the expected value of the highest Hamming Similarity score between $h^*$ and the hashes of all sampled images: 
    \begin{equation}
        s = \expvsub{\max_{x_i}\;{\mathrm{sim}(h^*, \nha{x_i})}}{\{x_1,\dots,x_n\} \sim X}.
    \end{equation}
    Given $n$ queries to the hashing oracle, it should be hard for an adversary to consistently generate near collisions:
    \begin{equation}
        \expvsub{\mathrm{sim}(\nha{x}, h^*)}{x} \leq s + \negl.
    \end{equation}
    The ability for an adversary to do so would suggest that $\nha{x}$ leaks information about $x$.
\end{property}

\begin{figure*}[t]
  \centering
  \includegraphics[keepaspectratio, width=.8\textwidth]{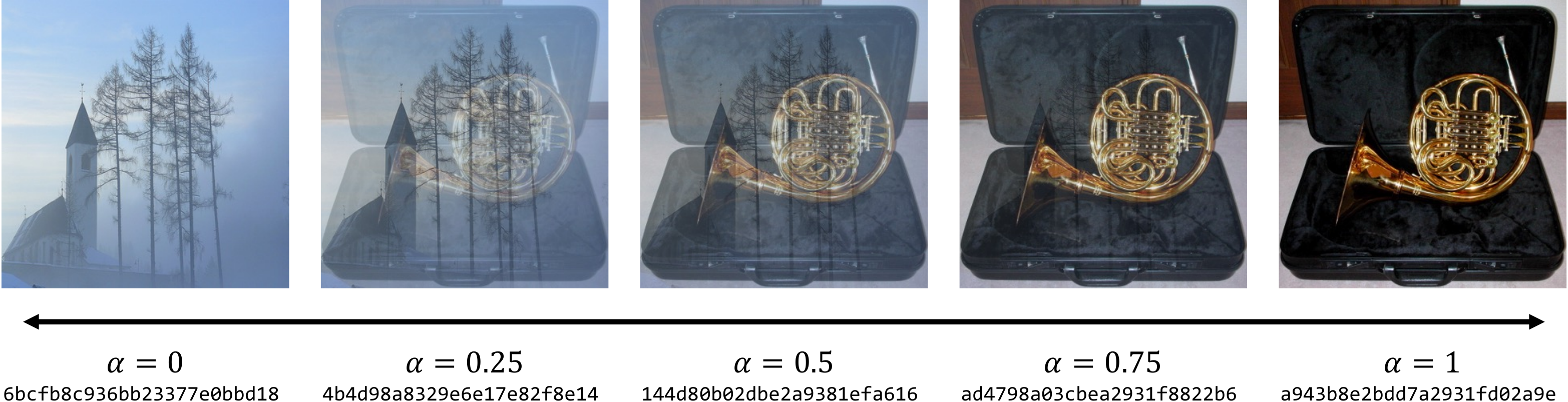}
  \caption{Images produced by interpolation of a face and a tree, with parameter $\alpha$ denoting the proportion of the first image.}
  \label{fig:Interp-graphic}
\end{figure*}

Note that our security goals for \nh are considerably stricter than those of other perceptual hashes \cite{msft-dna, fb-pdq}, particularly in terms of privacy (Properties \ref{prop:uniform-hashing}, \ref{prop:strong-non-invert}). This is because \nh is used in a \textit{client-side scanning} (CSS) scheme \cite{apple-psi} on users' devices, where potential false positives or privacy leaks would significantly damage trust between consumers and providers.

\subsection{Functional Description of \nh} \label{subsec:nh-description}

\nh hashes are generated in two steps. First, a MobileNet-based \cite{howard2017mobilenets} convolutional neural network generates a feature vector describing the image using 128 floating-point numbers. This vector is then multiplied by a randomized hashing matrix; the result is then thresholded using a step function to produce the final 96-bit binary hash. See Figure \ref{fig:NH-Diag} for a visual depiction.

Note that the hashing matrix can be interpreted as containing ninety-six random 128-dimensional hyperplanes. To generate the hash, the 128-dim feature vector is compared with each of these hyperplanes; if the feature vector is on the positive side of the hyperplane, the bit is 1, else 0.

\subsection{Related Work in Attacking \nh} \label{subsec:related-work}






In the realm of intervention-free attacks, \citet{nh-collisions-blog, nh-collisions-github} discovered naturally-occurring \nh collisions by mining publicly-available images that were not artificially modified in any way. This demonstrates a failure of Property \ref{prop:uniform-hashing}.


Because \nh relies on a neural network whose weights have been extracted and made public \cite{nh-repro-github}, a variety of gradient-based attacks have also been proposed \cite{breaking-nh-struppek}. These white-box attacks rely on the susceptibility of neural networks to small adversarial perturbations \cite{goodfellow2014explaining}, allowing malicious actors to both evade detection (violating Property \ref{prop:neighborhood}) and generate collisions (violating Properties \ref{prop:weak-non-invert}, \ref{prop:strong-non-invert}) using small amounts of adversarial noise.


Even more interesting, however, would be the existence of strong \textit{black-box} attacks; it is generally accepted that adversarial examples are an unavoidable phenomenon in neural networks.
\citet{breaking-nh-struppek} makes initial headway into this line of inquiry, showing that basic image transformations including translation, rotation, flipping, cropping, and brightness/contrast changes can help evade detection (violating Property \ref{prop:neighborhood}). The other security properties remain underexplored in the black-box setting.

In summary, the existing literature around gradient-based white-box attacks is fairly well-developed, but we lack a sufficiently deep understanding of \nh's behavior to develop strong black-box attacks.

%% file: report/attacks.tex
\section{\nh's Approximate Linearity} \label{sec:nh-linearity}


Imagine that we have two images $x_1, x_2$ where $d(x_1, x_2) > \epsilon$. By Property \ref{prop:uniform-hashing}, they should initially have random-looking hashes on average, i.e. $\expv{\mathrm{sim}(\nha{x_1}, \nha{x_2})} \leq 0.5 + \negl$. But what happens as we gradually move $x_2$ towards $x_1$?
In an idealized \nh, these hashes should remain random-looking until $d(x_1, x_2^\prime) \leq \epsilon$, at which point the hashes should match exactly.

A natural strategy for analyzing the gradual movement of images toward each other is \textit{image interpolation}. To the best of our knowledge, this has not yet been studied in the context of \nh.

More formally, consider taking two source images $x_1$ and $x_2$ and linearly interpolating them with respect to some parameter $\alpha$. The resulting image is $I_{\alpha} = \alpha x_1 + (1-\alpha) x_2$.
Figure \ref{fig:Interp-graphic} provides a visualization of interpolation where $\alpha$ is varied from 0 to 1 between an image of a French horn ($x_1$) and an image of a church ($x_2$).

\begin{figure}[ht]
  \centering
  \includegraphics[keepaspectratio, width=0.8\columnwidth]{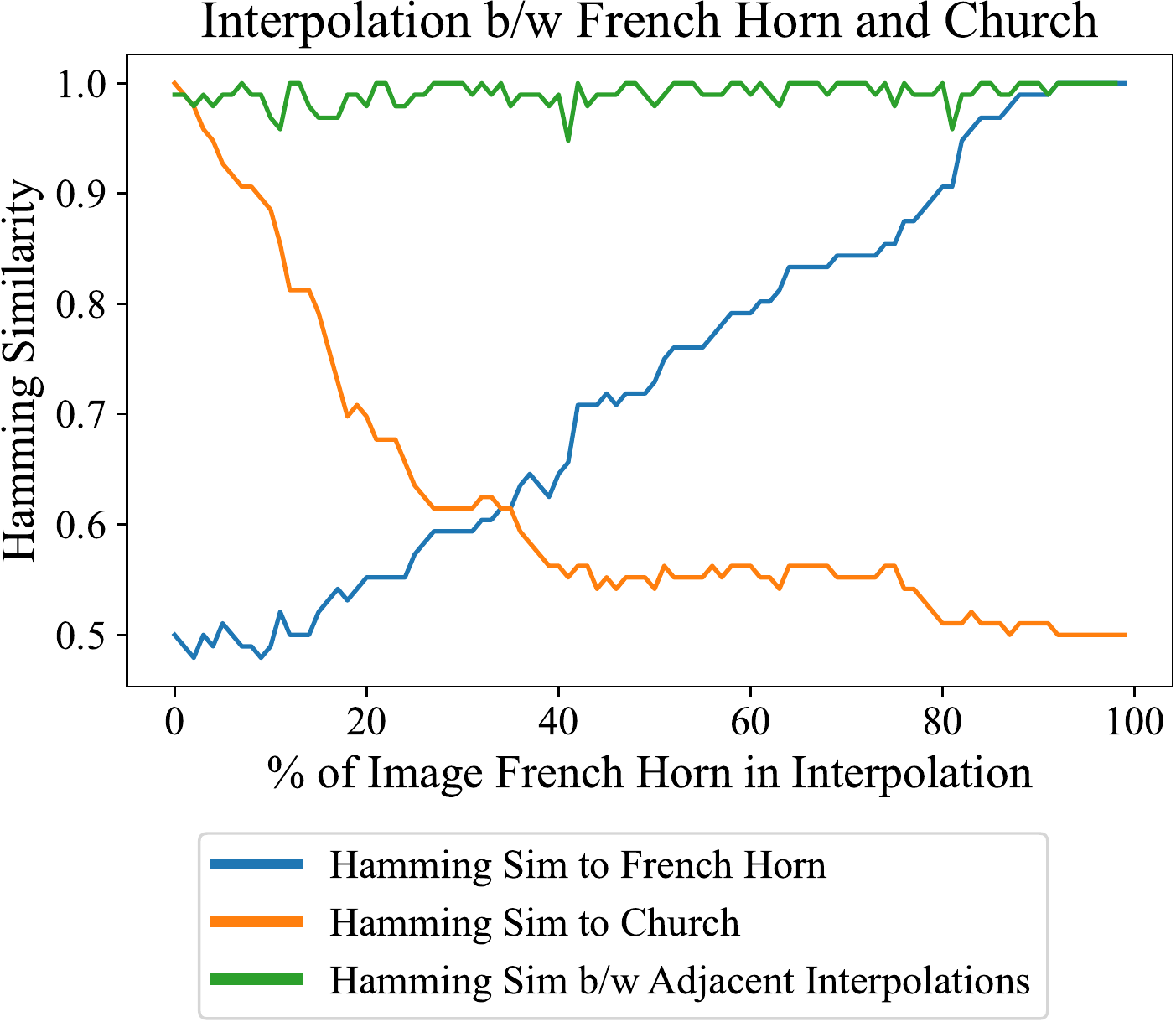}
  \caption{\textbf{Approximate Piecewise Linearity of \nh}. When interpolating between two images, we observe violations of \nh's security criteria (Property \ref{prop:uniform-hashing}). Hashes of semantically-unrelated images do \textit{not} look random.}
  \label{fig:Interp-graph-one}
\end{figure}

To quantitatively analyze the behavior of \nh as $x_1, x_2$ are interpolated, we plot the Hamming similarity between $I_\alpha$ and both $x_1, x_2$ for all $\alpha$; Figure \ref{fig:Interp-graph-one} displays the results.
Strikingly, as $\alpha$ moves from 0 to 1, the Hamming similarity between $I_{\alpha}$ and the original images $x_1, x_2$ changes approximately \textit{piecewise linearly}.

This provides strong evidence that \nh fails to meet its privacy requirements.
Specifically, by Property \ref{prop:uniform-hashing}, we would expect (i) the orange and blue lines to remain at roughly $0.5$ in the middle region, where $I_\alpha$ is semantically-identical to neither $x_1$ nor $x_2$, and (ii) the green line to oscillate between $1$ and $0.5$. Empirically, we observe failures of both expectations.


\begin{figure}[ht]
  \centering
  \includegraphics[keepaspectratio, width=0.8\columnwidth]{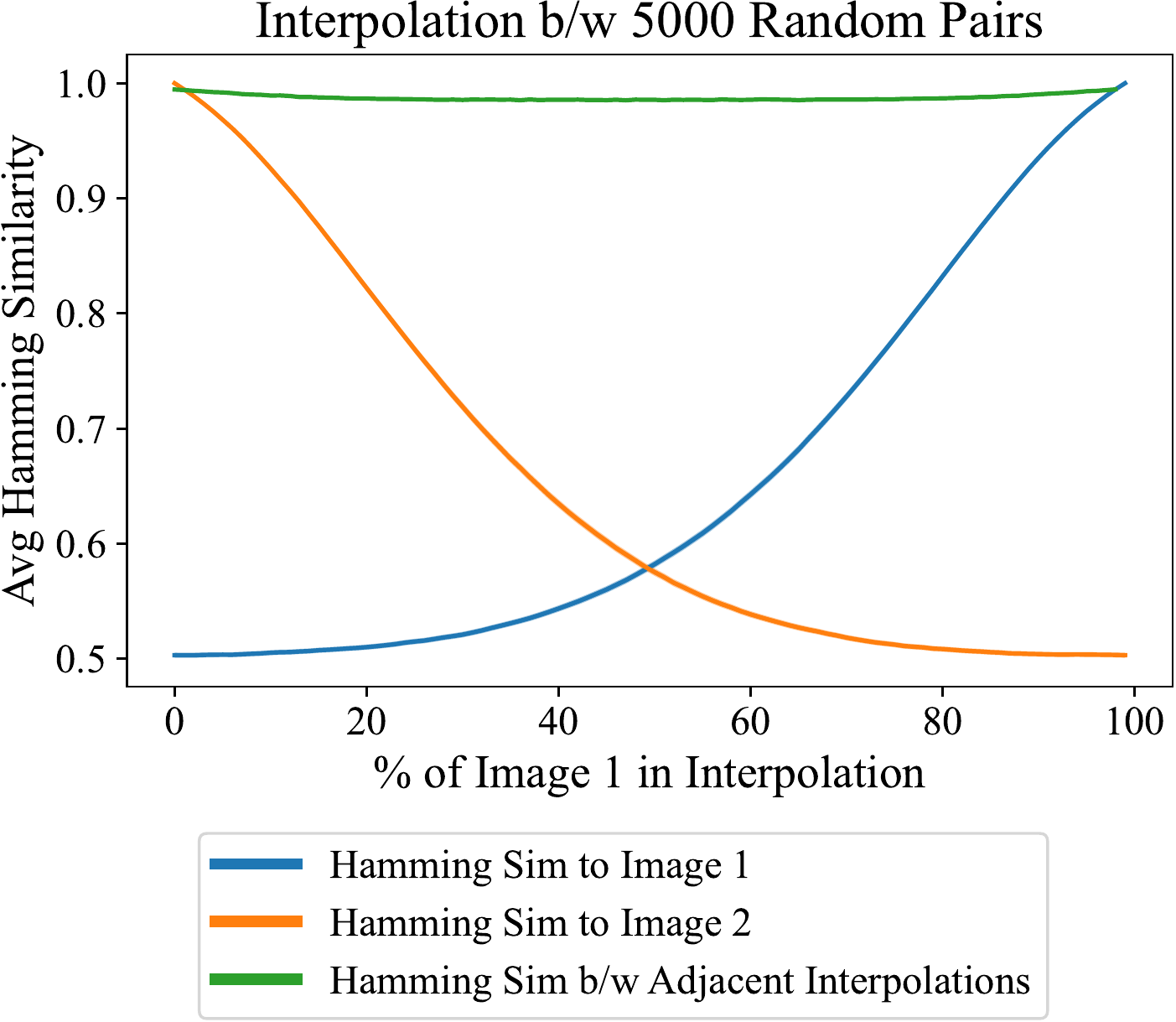}
  \caption{\textbf{Averaged Interpolation Behavior}. We average interpolation similarities over a large set of randomly sampled ImageNette pairs, plotted with $99\%$ confidence intervals. The approximate linearity of \nh is systematic.}
  \label{fig:Interp-graph-many}
\end{figure}

Furthermore, this approximate piecewise linearity is \textit{systematic}. In Figure \ref{fig:Interp-graph-many}, we average Hamming similarities over a large set of ImageNette \footnote{ImageNette is a subset of ImageNet containing roughly 1000 examples from 10 classes. Note that since \nh is meant to be a generalized perceptual hashing algorithm, it is reasonable to use any dataset for our experiments.} \cite{imagenette} pairs, finding the same pattern with tight $99\%$ confidence intervals.


These results are significant not only because they demonstrate the violation of several security properties. Perhaps more importantly, they clarify a key operational behavior of \nh, approximate linearity, which enables us to make fairly reliable inferences about the hashes of the interpolations between a set of images.

\section{Attacks} \label{sec:attacks}

The approximate linearity of \nh lends itself naturally to several attacks. Firstly, the interpolation graphs (Figures \ref{fig:Interp-graph-one}, \ref{fig:Interp-graph-many}) suggest the possibility of an \textbf{evasion attack} where, given a target image $x$, we can compute a semantically similar image $x'$ by adding a small percentage of another image, such that  $\nha{x} \ne \nha{x'}$. This would break the neighborhood property.

The predictability of \nh's hashing behavior also suggests that we might be able to \textbf{generate near-collisions} where, given any target image $x$, we can \textit{on average} generate an image $x'$ that nearly collides with but is not semantically similar to $x$: $\expvsub{\simfn(\nha{x}, \nha{x'})}{x} \gg 0.5 + \negl$. This breaks uniform hashing and strong non-invertibility.

Finally, we hypothesize that \nh \textbf{leaks information}. Given an image $x$ which belongs to one of $n$ known image classes $\{C_1, C_2, ..., C_n \}$, we can predict target class $C_t$ with probability greater than $\frac{1}{n} + \mathrm{negl}$. This would break the uniform hashing property.

In the following subsections, we demonstrate that all attacks can be realized using interpolation-based algorithms.

\subsection{Evasion Attack} \label{subsubsec:interp-evasion}

\textbf{Attack Description}. Given a source image $x$ and an arbitrary semantically-unrelated image $x_0$, we produce a semantically-similar image $x' = \alpha^* x + (1-\alpha^*) x_0$ such that $\nha{x} \ne \nha{x'}$. $\alpha^*$ is chosen as the largest $\alpha$ for which the hashes differ:
\begin{equation} \label{eq:evasion-alpha-star}
    \alpha^* = \max \bigg\{ \alpha \mid \nha{x} \neq \nha{\alpha x + (1-\alpha) x_o} \bigg\}
\end{equation}
Intuitively, this selects for the interpolation that contains the maximum possible percentage of $x$.
 
 
\textbf{Attack Results}.
We consider an attack ``successful'' iff (i) $\nha{x} \neq \nha{x^\prime}$ and (ii) $x$ and $x^\prime$ contain the same semantic information. To formalize the latter concept, we adopt the Structural Similarity Index Measure (SSIM) metric, as used in \citet{breaking-nh-struppek}:
\begin{equation} \label{eq:ssim}
    \ssim(x, y) = \frac{(2\mu_x\mu_y+c_1)(2\sigma_{xy} + c_2)}{(\mu_x^2 + \mu_y^2 + c_1)(\sigma_x^2 + \sigma_y^2 + c_2)},
\end{equation}
where $x$ and $y$ are images with means $\mu_x, \mu_y$, variances $\sigma_x^2, \sigma_y^2$, and covariance $\sigma_{xy}$. Moreover, $c_1=0.01, c_2=0.03$ for numerical stability.
The intuition is that, by accounting for mean, variance, and covariance in the pixel space, SSIM focuses on \textit{structure} better than Mean-Squared Error (MSE), for example, which can be distracted by noise.\footnote{Note that SSIM is \textit{not} robust under various transformations such as rotation or translation, but this is not a concern because the interpolation attack is executed with a pixelwise sum.}

\begin{figure}[ht]
  \centering
  \includegraphics[keepaspectratio, width=1\columnwidth]{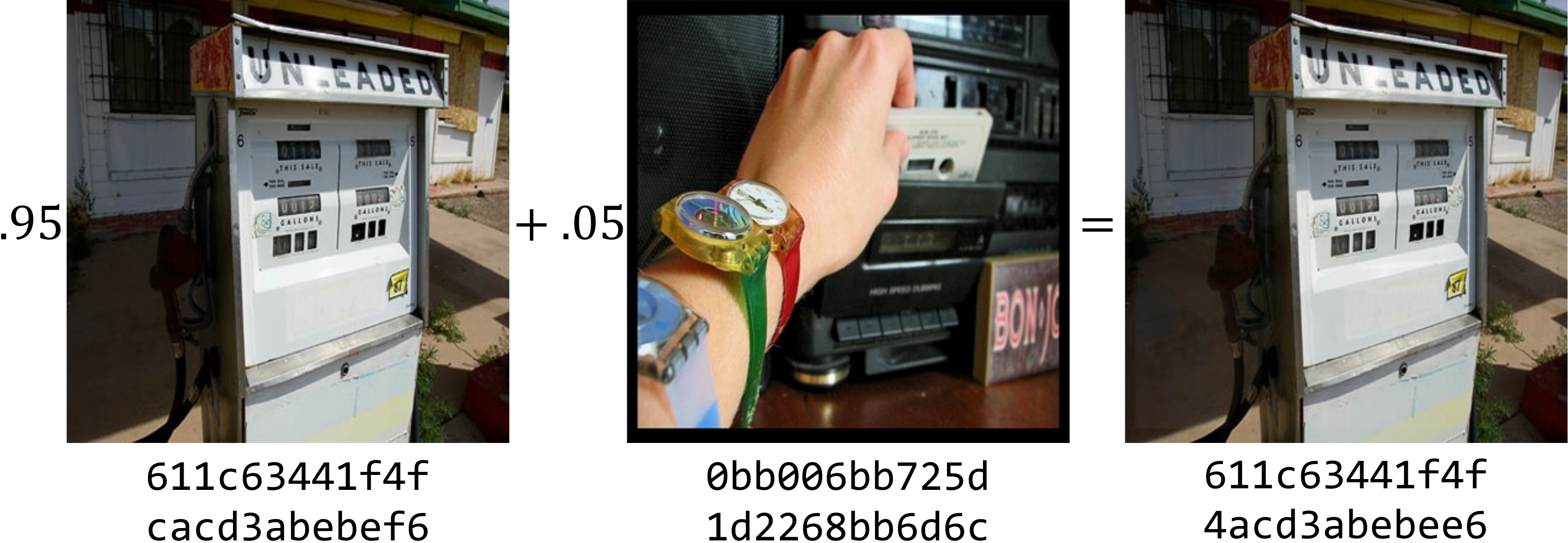}
  \caption{\textbf{Qualitative Example of Evasion}. Interpolating between the gas pump and cassette player ($\alpha^* = 0.95$, as computed using Eqn. \ref{eq:evasion-alpha-star}) results in an image that is semantically indistinguishable from the original gas pump, but has a different hash.}
  \label{fig:evasion-example}
\end{figure}

We carry out a systematic evaluation by randomly sampling 10,000 pairs of images from ImageNette, designating one of each pair to be $x$ and the other to be $x_0$. On average, the attack achieves \textbf{100\% efficacy} with an
\textbf{SSIM of 0.981} ($\pm 5\times 10^{-4}$ for a 95\% confidence interval).
Figure \ref{fig:evasion-example} shows a representative example of the attack. The interpolated image is hardly distinguishable from the original gas pump picture, but its hash is different in multiple bit positions.


Our interpolation-based evasion attack has several advantages over existing methods. In particular, transformations such as rotation, cropping, translation, reflection, and brightness/contrast adjustment \cite{breaking-nh-struppek} \textit{do} preserve semantic content but nonetheless result in significant changes to the appearance of images. By comparison, our interpolation attacks cause minimal visual impact and involve less guesswork than performing arbitrary transformations. A limitation, however, is that the efficacy of the interpolation attack is somewhat dependent on a good choice of $x_0$. We show that, on average, the method is highly robust, but there are a few outlier cases where a poor choice of $x_0$ causes SSIM to drop below 0.7.

\subsection{Near-Collision Generation Attack} \label{subsec:interp-collision}

\textbf{Attack Description}. Consider a database of images $X = \{x_1, x_2, .., x_k\}$. Let $P$ be a parameter space describing all possible interpolations of images in $X$. Thus, each element $p\in P$ represents an interpolated image $I_p$ where
\begin{equation} \label{eq:interp-over-p}
    I_p = \sum_{i=1}^k p_i x_i,
\end{equation}
and $p_i \in [-1, 1]$ and $\sum_{i=1}^k |p_i| = 1$.
Then, given a target hash $h^*$, we employ a genetic algorithm \cite{whitley1994genetic} to search $P$ for an interpolated image whose hash maximizes Hamming similarity with $h^*$. Formally, we optimize:
\begin{equation}
    \max_{p \in P} \, {\simfn(\nha{I_p}, h^*)}
\end{equation}
using the Hamming similarity score defined in Eqn. \ref{eq:hamming-sim} and with $I_p$ as defined in Eqn. \ref{eq:interp-over-p}.

Crucially, our algorithm employs two genetic operators that rely heavily on the properties of interpolated images explored in Section \ref{sec:nh-linearity}.
\vspace{-5pt}
\begin{description}
    \item [Crossover Operator] Given $p, q \in P$ and $\alpha \in [0, 1]$. Compute a new crossed-over parameter $r = \alpha p + (1-\alpha)q$. Note that $r$ is normalized after interpolation such that $I_r \approx \alpha I_p + (1-\alpha)I_q$. By the interpolation property, with high probability, $\simfn(\nha{I_r}, h^*)$ will be between $\simfn(\nha{I_p}, h^*)$ and $\simfn(\nha{I_q}, h^*)$, and with some nontrivial probability it will be greater.
    \item [Mutation Operator] Given $p \in P$, index $m \in \{1, \dots, k\}$ and $\alpha \in [-0.05, 0.05]$. Compute a new mutated parameter $r$ by mutating $p_m$ by $\alpha$. More formally, $\forall i \ne m. \; r_i = p_i$, and $r_m = p_m + \alpha$. By our interpolation property, such a mutation should only slightly change the Hamming similarity. Additionally, with some nontrivial probability, $\simfn(\nha{I_r}, h^*) > \simfn(\nha{I_p}, h^*).$ As with the crossover operator, note that $r$ is normalized after mutation.
\end{description}

\vspace{-5pt}

We use a population size of $100$ decreasing exponentially to $10$ at a rate of $0.97$. Our algorithm runs for $50$ iterations, each time generating $20$ children. Each child is equally likely to be formed as a product of a Crossover or Mutation operator, and all other parameters ($\alpha, m, p, q$) are selected uniformly at random from the appropriate ranges.

Our database $X$ consists of $250$ images, $25$ training images from each of the $10$ ImageNette classes \cite{imagenette}. We test our algorithm on the hashes of $150$ images, $15$ validation images from each of the same $10$ classes. To improve performance, our hashes are encoded as $\{-1, 1\}^{96}$ vectors rather than the standard $\{0, 1\}^{96}$.\footnote{Linear combinations of the value $0$ are not meaningful.}

Note that the approximate linearity of \nh is crucial in enabling this attack, because it narrows the search space from all possible images (which is unimaginably large) to $P$, the set of interpolated images within a fixed dataset. This mitigates the combinatorial explosion and makes the problem tractable.



\begin{figure}[ht]
  \centering
  \includegraphics[keepaspectratio, width=0.8\columnwidth]{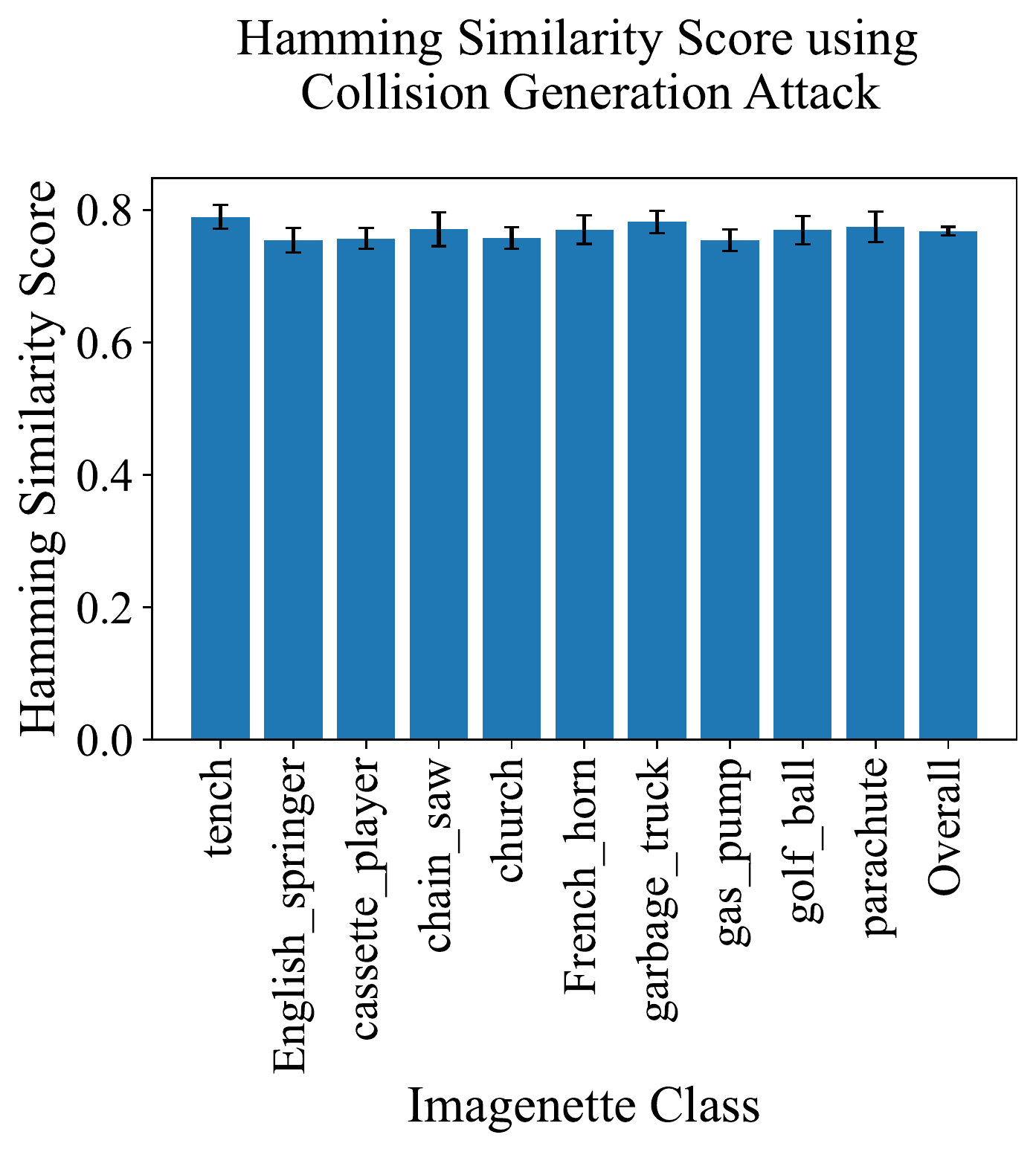}
  \caption{\textbf{Near Collision Generation with Genetic Algorithm}. Hamming similarity score of images generated by the genetic algo, averaged across $150$ target hashes: $15$ per class $\times$ 10 classes.}
  \label{fig:Interp-Collision}
\end{figure}

\begin{figure*}[ht]
  \centering
  \includegraphics[keepaspectratio, width=0.8\textwidth]{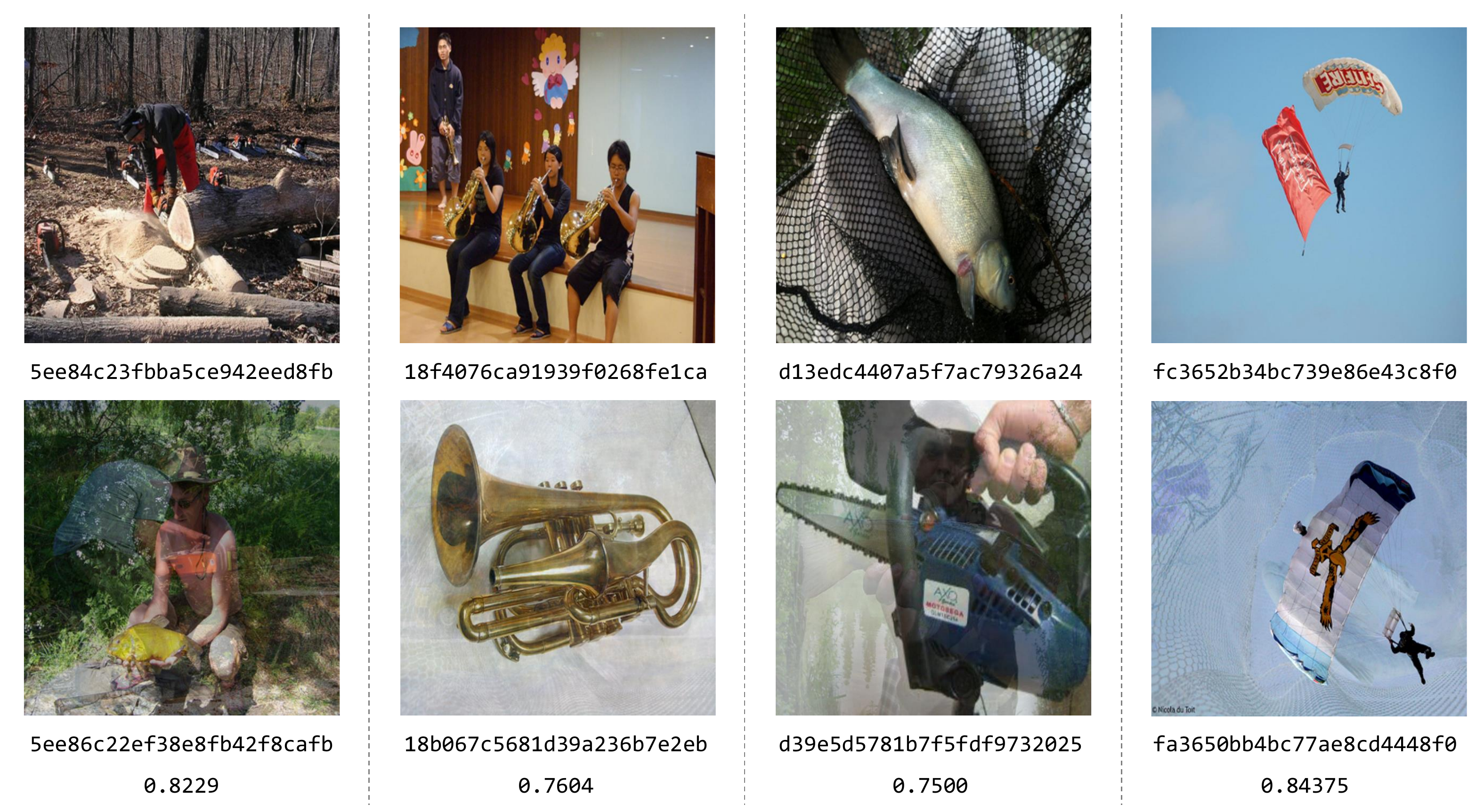}
  \caption{\textbf{Near Collisions Created by the Genetic Algorithm}. The first row contains the source image and its hash (the target hash). The second row contains the interpolated image optimized by the genetic algorithm, its hash, and the Hamming similarity between the interpolated image's hash and the target hash.}
  \label{fig:Interp-Collision-Vis}
\end{figure*}

\textbf{Attack Results}. 

Our results are shown in Figure \ref{fig:Interp-Collision}. While we are not directly able to find perfectly colliding images, the average Hamming similarities of generated images are about \textit{six} standard deviations above the mean of $0.5$,\footnote{Viewing hashes as a binomial RV, $\sigma = \sqrt{npq} \approx 0.05$.} indicating that the genetic algorithm performs much better than random generation and thus violates strong non-invertibility (Property \ref{prop:strong-non-invert}). This conclusion is further reinforced in Section \ref{sec:improvements} where we perform the same attack after improving the security of the \nh system (see Figure \ref{fig:Interp-Col-Sha}).


Qualitatively, it is also interesting to see that the genetic algorithm's generations often share semantic similarities in color and shape with the source image, from which the target hash is derived (Figure \ref{fig:Interp-Collision-Vis}).


\subsection{Information Extraction Attack} \label{subsec:info-extraction}

Recall the goal of information extraction from Section \ref{sec:attacks}, where the goal is to predict the class of an image $x$ from a set of $n$ classes with probability greater than $\frac{1}{n} + \mathrm{negl}$ given only its hash, $h^*$. We also assume access to a database of images $\{x_1, x_2, .., x_k\}$ with known hashes $\{h_1, h_2, .., h_k\}$ and known class labels $\{c_1, c_2, .., c_k\}$. 

\textbf{Attack Description}. Let $\{p_1, p_2, .., p_k\} \in [-1, 1]^k$ be parameters such that $\sum_{i=1}^k |p_i| = 1$.
Our attack is simple: first, minimize the following objective using gradient decent over parameters $\{p_1, p_2, .., p_k\}$
\begin{equation} \label{eq:obj-information-extraction}
    L(p_1, p_2, .., p_k, h^*) = \left(\sum_{i=1}^k p_ih_i - h^*\right)^2 - \sum_{i=1}^k p_i \log(p_i).
\end{equation}

Then, compute the class $C$ with the most support:
\begin{equation}
    \max_C {\sum_{i \text{ s.t. } c_i = C } |p_i|}
\end{equation}

Note that this attack does not require model access except when labeling the dataset, as it only utilizes the hashes of the dataset images $h_i$, and class labels $c_i$.

To understand why this attack is reasonable, consider the objective function $L(\cdot)$ from Eqn. \ref{eq:obj-information-extraction}. Imagine if we could find a setting of parameters $\{p_1, p_2, .., p_k\}$ such that $\sum_{i=1}^k p_ih_i = h^*$.
The interpolation property of the \nh model would imply that $\smash{\nha{\sum_{i=1}^k p_ix_i} \approx h^*}$.
This motivates the first term of $L(\cdot)$.\footnote{One might wonder why we chose not to use a similar objective in Section \ref{subsec:interp-collision} while generating near collisions. We find that, because the space of solutions to Eqn. \ref{eq:obj-information-extraction} is fairly large, it is difficult to consistently generate reasonable-looking in-distribution images that resulted in hash collisions. However, we find that it \textit{does} leak a significant amount of class information.} The second term simply aims to reduce entropy or, in other words, use the fewest number of parameters possible to achieve the objective.

Some additional information about our approach is as follows. Our database $X$ consists of $500$ images, $50$ training images from each of the $10$ ImageNette classes \cite{imagenette}. We tested our algorithm on the hashes of $1000$ images, $100$ validation images from each of the same $10$ classes. Our algorithm runs for $25$ epochs, $100$ steps per epoch, and with a learning rate of $2\cdot 10^{-5}$. To improve performance, our hashes are encoded as $\{-1, 1\}^{96}$ vectors rather than the standard $\{0, 1\}^{96}$.

\begin{figure}[ht]
  \centering
  \includegraphics[keepaspectratio, width=0.8\columnwidth]{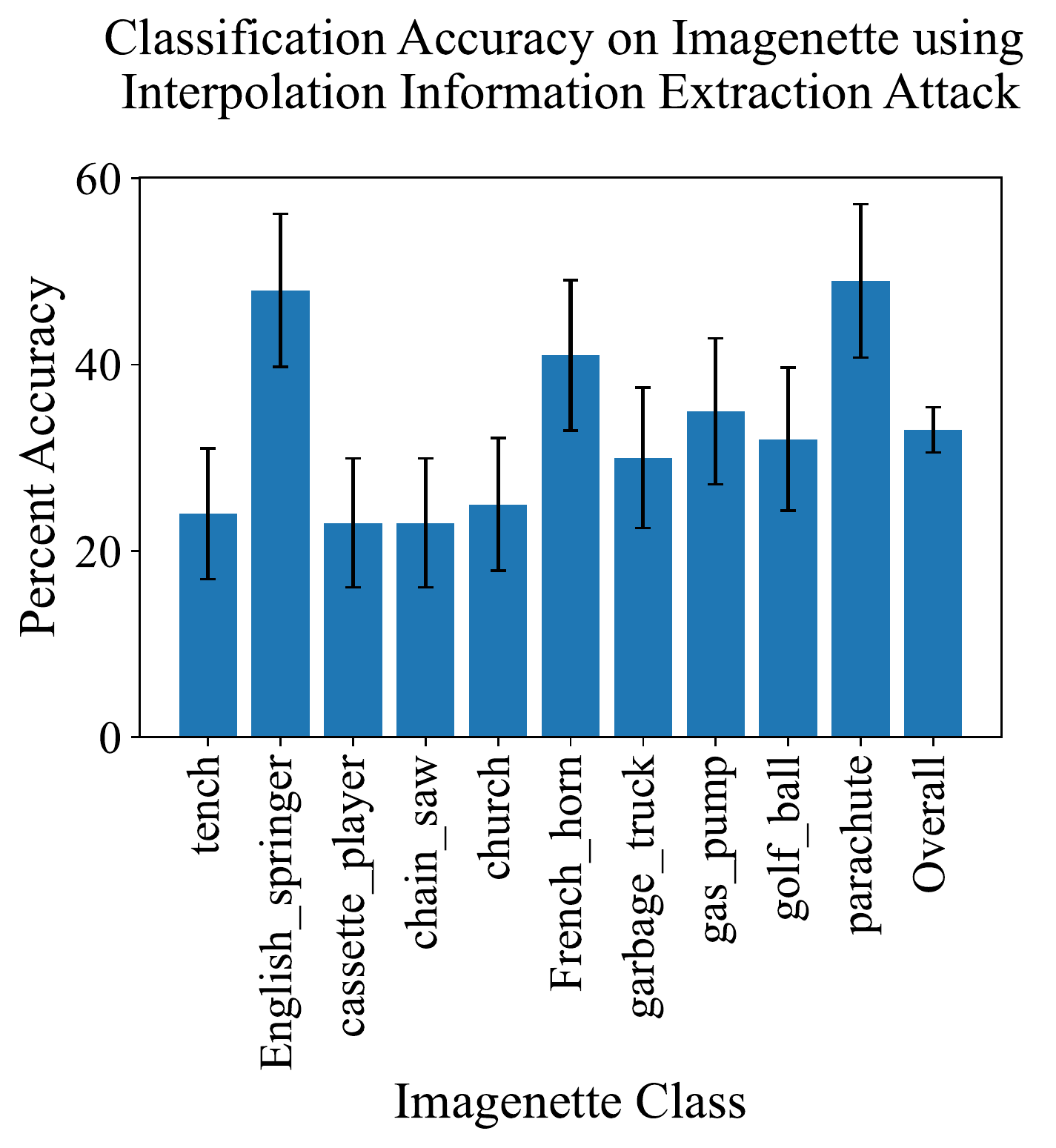}
  \caption{\textbf{Classification Accuracy Shows Significant Information Leakage}. Classification accuracy of ImageNette images from just their hash using an Interpolation Information Extraction Attack. Results averaged across $1000$ target hashes: $100$ per class over $10$ classes.}
  \label{fig:Interp-Info-Extract}
\end{figure}

\textbf{Attack Results}. Our results, seen in Figure \ref{fig:Interp-Info-Extract} show that \nh hashes leak a  significant amount of information about the semantic contents of images they are derived from. This is a clear violation of the Uniform Hashing Property, Property \ref{prop:uniform-hashing}. It is interesting that classification accuracy (and therefore information leakage) seems to be quite class dependent. As one might expect, classes containing images with less clutter and clearer subjects tend to have more information leakage.

%% file: report/system-improvements.tex
\section{Improving \nh: SHA-at-the-End} \label{sec:improvements}





All black-box interpolation attacks presented in Section \ref{sec:attacks} all rely on the approximate linearity of $\nha{x}$. We aim to nullify this property.

\newcommand{\sha}[1]{\mathrm{SHA}\left(#1\right)}

\textbf{Method Description}. To address the undesirable linearity of \nh, we propose a simple yet effective extension: adding a SHA-256 block at the very end of the computation. Revisiting the security definitions from Section \ref{sec:prelim}, we find that this strengthens the uniform hashing and noninvertibility properties while maintaining the neighborhood property: if $\nha{x} = \nha{x^\prime}$, then it must be true that $\sha{\nha{x}} = \sha{\nha{x^\prime}}$. Moreover, if $\nha{x} \neq \nha{x^\prime}$, then under the Random Oracle Model (ROM) their hashes will (i) look uniformly random and (ii) be non-invertible.

\begin{figure}[ht]
  \centering
  \includegraphics[keepaspectratio, width=0.8\columnwidth]{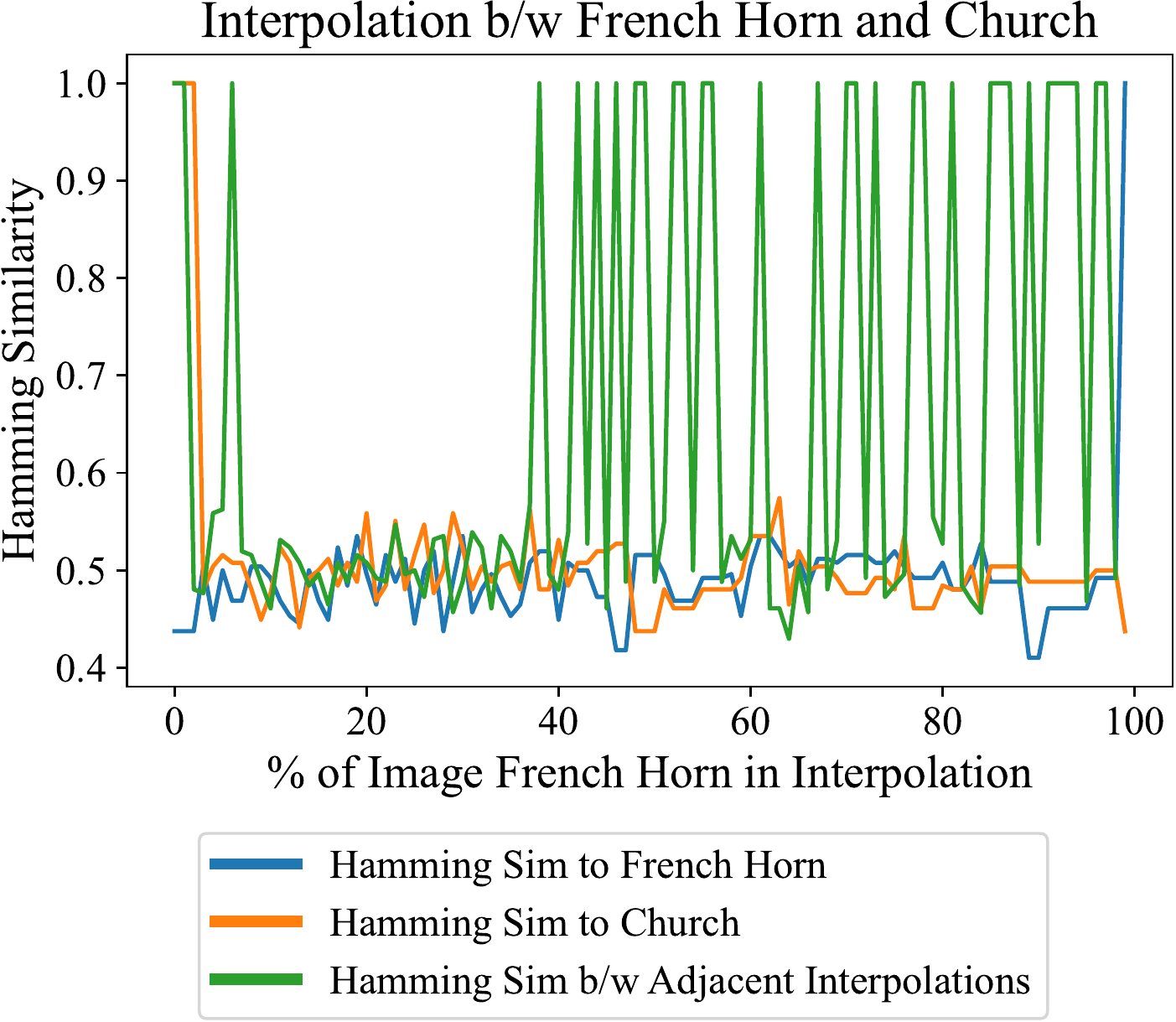}
  \caption{\textbf{Interpolation Results using SHA-at-the-End}. As expected, SHA breaks the linearity of \nh (compare to Figure \ref{fig:Interp-graph-one}).}
  \label{fig:interp-sha}
\end{figure}

\begin{figure}[ht]
  \centering
  \includegraphics[keepaspectratio, width=0.8\columnwidth]{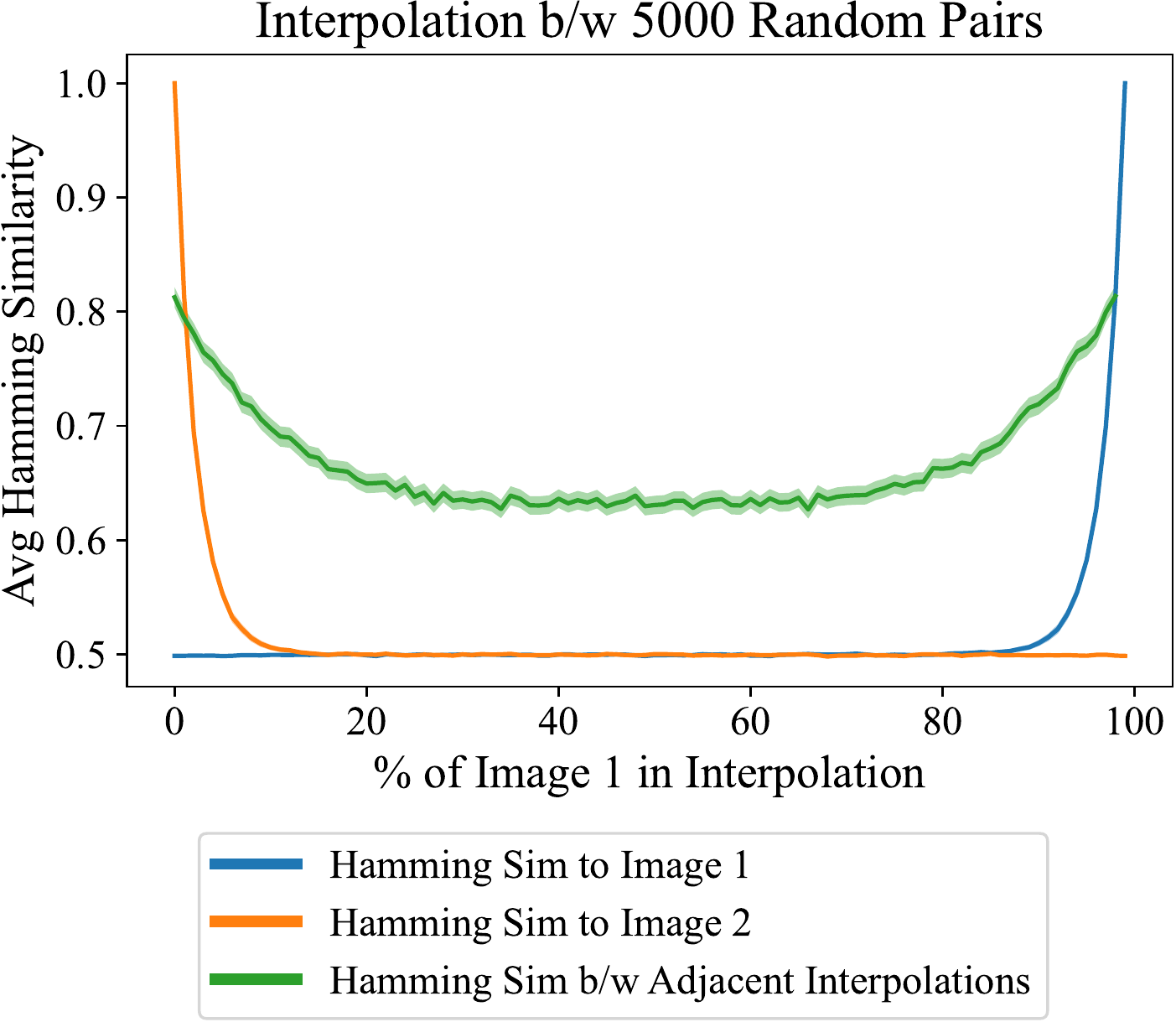}
  \caption{\textbf{Averaged Interpolation Under SHA}. Here we demonstrate that the impact of SHA is systematic, and the violation of security properties is alleviated (compare to Figure \ref{fig:Interp-graph-many}).}
  \label{fig:interp-many-sha}
\end{figure}



\textbf{Method Results}.
First, we check whether the interpolation graph behaves as expected. As seen in Figure \ref{fig:interp-sha}, the Hamming similarities hover around 0.5 (except at the borders), and the similarity between adjacent interpolations oscillates between 1 and $\approx 0.5$. Figure \ref{fig:interp-many-sha} demonstrates that these trends are systematic. This suggests that SHA-at-the-end successfully breaks approximate linearity.

\begin{figure}[ht]
  \centering
  \includegraphics[keepaspectratio, width=0.8\columnwidth]{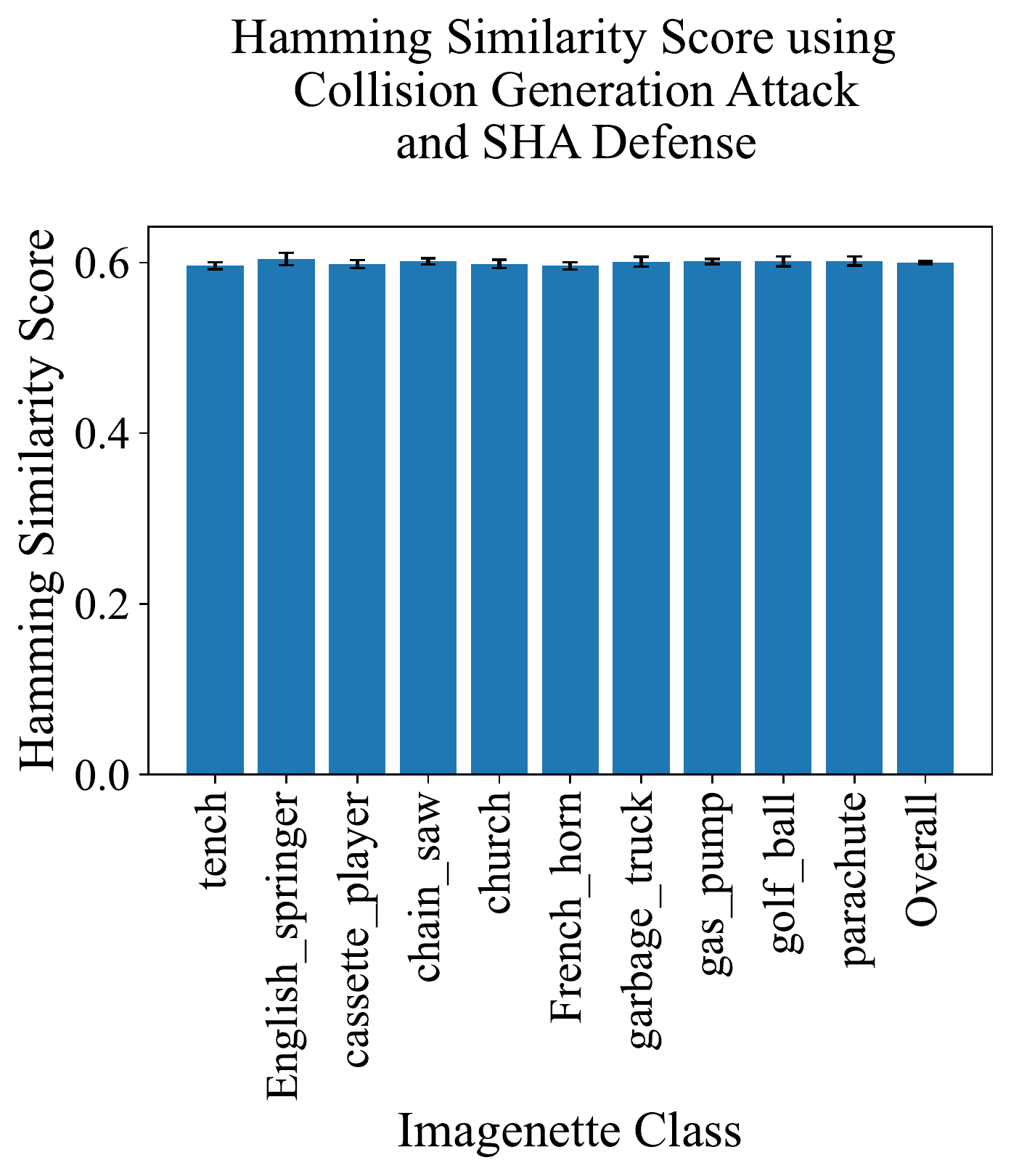}
  \caption{\textbf{SHA Hinders Collision Generation with Genetic Algorithm}. Hamming similarity score of images generated by the genetic algorithm averaged across $150$ SHA-ed target hashes: $15$ per class over $10$ classes (compare to Figure \ref{fig:Interp-Collision}).}
  \label{fig:Interp-Col-Sha}
\end{figure}

\begin{figure}[ht]
  \centering
  \includegraphics[keepaspectratio, width=0.8\columnwidth]{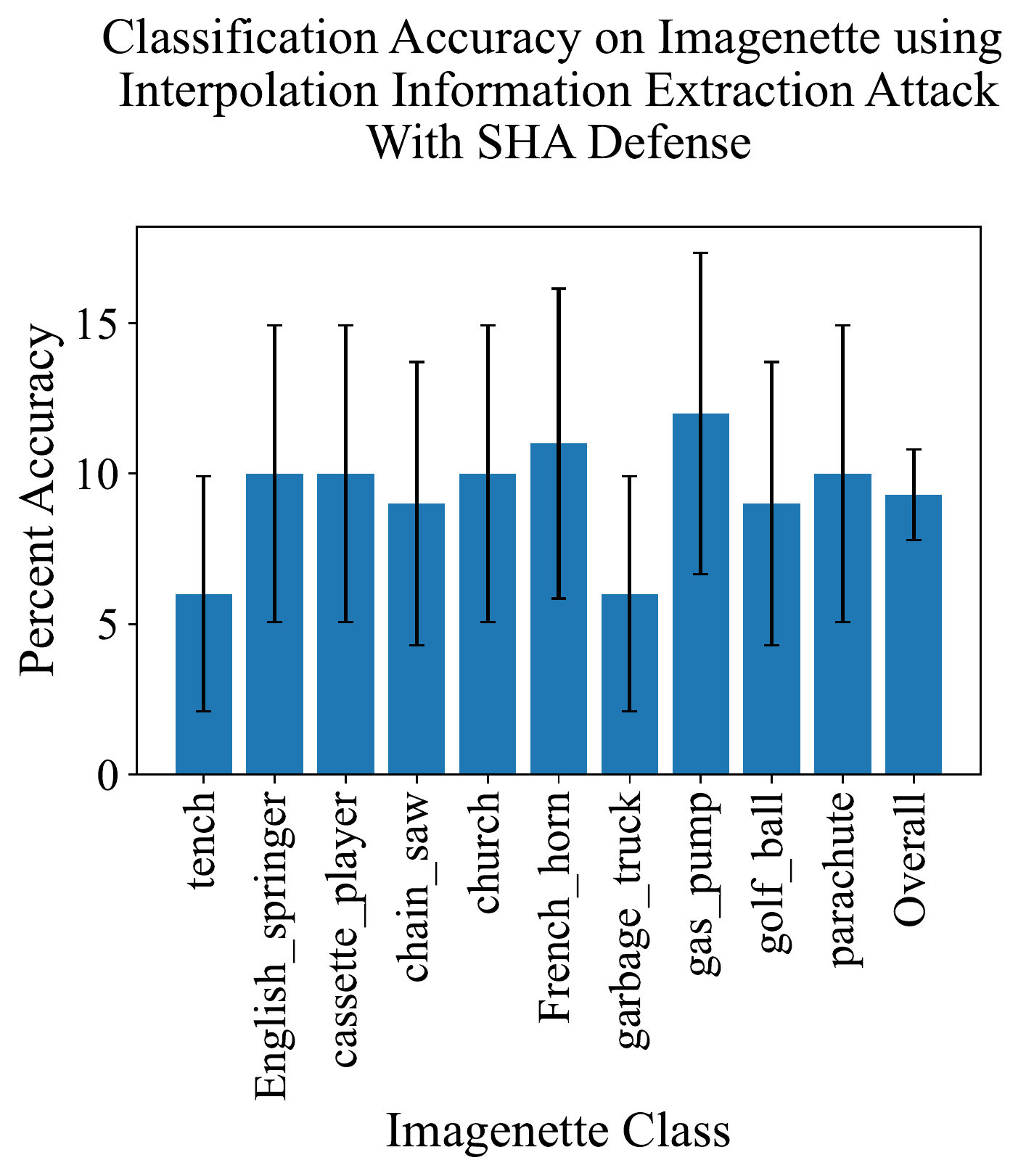}
  \caption{\textbf{SHA Disables Information Leakage}. Imagenette classification accuracy from SHA-ed hashes using an Interpolation Information Extraction Attack. Results averaged across $1000$ target hashes: $100$ per class over $10$ classes (compare to Figure \ref{fig:Interp-Info-Extract}).}
  \label{fig:Interp-Info-Extract-Sha}
\end{figure}

We further these conclusions experimentally by re-attempting the same interpolation attacks for collision generation (Section \ref{subsec:interp-collision}) and information extraction (Section \ref{subsec:info-extraction}) with the new system architecture.\footnote{Note that we do not attempt to defend against the evasion attack (Section \ref{subsubsec:interp-evasion}), as it will clearly work just as well. While the evasion attack does rely on the interpolation property, it is ultimately determined by \nh's neighborhood property, which our proposed change has no impact on.}
From Figure \ref{fig:Interp-Col-Sha}, we see that the ability to generate near-collisions is significantly hindered by the SHA block. The algorithm still consistently produces images with Hamming similarities greater than $0.5$, but this is reasonable; the expected similarity of the \textit{maximum} of many randomly-sampled hashes is greater than the expectation of just one hash. From Figure \ref{fig:Interp-Info-Extract-Sha}, we also see that SHA-at-the-end completely disables information leakage, and the overall classification accuracy is approximately 10\%, as expected. We conclude that SHA provides a simple defense mechanism against both these attacks.

%% file: report/conclusion.tex
\section{Conclusion}

In this paper, we have proposed and analyzed a variety of black-box attacks that expose vulnerabilities of Apple's \nh algorithm. In particular, our black box attacks suggest that \nh is fundamentally flawed \textit{aside} from the existence of gradient-based adversarial examples; this is a much stronger statement than the white-box case. At the core of these attacks is our discovery that \nh's outputs change approximately linearly with respect to its inputs. Furthermore, we suggest a simple modification and demonstrate that it nullifies the piecewise linear property while successfully safeguarding against some of our proposed attacks.

This goes to show the immense difficulty of designing perceptual hashing algorithms that are both robustly effective and private. There remains much work to be done before such systems can be integrated into client-side scanning for real-world use.

\textbf{Acknowledgements}. We thank Ronald Rivest, Yael Kalai, Andres Fabrega, Kyle Hogan, Nithya Attaluri, and Albert Xing for their valuable assistance and feedback.

%% file: main.bbl
\begin{thebibliography}{17}
\providecommand{\natexlab}[1]{#1}
\providecommand{\url}[1]{\texttt{#1}}
\expandafter\ifx\csname urlstyle\endcsname\relax
  \providecommand{\doi}[1]{doi: #1}\else
  \providecommand{\doi}{doi: \begingroup \urlstyle{rm}\Url}\fi

\bibitem[Abelson et~al.(2021)Abelson, Anderson, Bellovin, Benaloh, Blaze,
  Callas, Diffie, Landau, Neumann, Rivest, Schiller, Schneier, Teague, and
  Troncoso]{abelson2021bugs}
Abelson, H., Anderson, R., Bellovin, S.~M., Benaloh, J., Blaze, M., Callas, J.,
  Diffie, W., Landau, S., Neumann, P.~G., Rivest, R.~L., Schiller, J.~I.,
  Schneier, B., Teague, V., and Troncoso, C.
\newblock Bugs in our pockets: The risks of client-side scanning, 2021.

\bibitem[Apple(2021)]{apple-nh}
Apple.
\newblock Child sexual assault material detection, 2021.
\newblock URL
  \url{https://www.apple.com/child-safety/pdf/CSAM_Detection_Technical_Summary.pdf}.

\bibitem[Athalye(2021)]{anish-github}
Athalye, A.
\newblock Neural hash collider, 2021.
\newblock URL \url{https://github.com/anishathalye/neural-hash-collider}.

\bibitem[Bhowmick et~al.(2021)Bhowmick, Boneh, and Myers]{apple-psi}
Bhowmick, A., Boneh, D., and Myers, S.
\newblock The apple psi system, 2021.
\newblock URL
  \url{https://www.apple.com/child-safety/pdf/Apple_PSI_System_Security_Protocol_and_Analysis.pdf}.

\bibitem[Chum et~al.(2008)Chum, Philbin, and
  Zisserman]{near-duplicate-im-detect}
Chum, O., Philbin, J., and Zisserman, A.
\newblock Near duplicate image detection: min-hash and tf-idf weighting.
\newblock \emph{BMVC 2008 - Proceedings of the British Machine Vision
  Conference 2008}, 01 2008.
\newblock \doi{10.5244/C.22.50}.

\bibitem[Di~Martino \& Sessa(2013)Di~Martino and Sessa]{fuzzy-transforms}
Di~Martino, F. and Sessa, S.
\newblock Image matching by using fuzzy transforms.
\newblock \emph{Advances in Fuzzy Systems}, 2013, 2013.
\newblock \doi{10.1155/2013/760704}.

\bibitem[Dwyer(2021)]{nh-collisions-blog}
Dwyer, B.
\newblock Imagenet contains naturally occurring neuralhash collisions, 2021.
\newblock URL \url{https://blog.roboflow.com/neuralhash-collision/}.

\bibitem[Facebook(2020)]{fb-pdq}
Facebook.
\newblock The pdq perceptual hash, 2020.
\newblock URL
  \url{https://github.com/facebook/ThreatExchange/blob/main/hashing/hashing.pdf}.

\bibitem[FastAI(2019)]{imagenette}
FastAI.
\newblock Imagenette: A subset of imagenet, 2019.
\newblock URL \url{https://github.com/fastai/imagenette}.

\bibitem[Goodfellow et~al.(2014)Goodfellow, Shlens, and
  Szegedy]{goodfellow2014explaining}
Goodfellow, I.~J., Shlens, J., and Szegedy, C.
\newblock Explaining and harnessing adversarial examples.
\newblock \emph{arXiv preprint arXiv:1412.6572}, 2014.

\bibitem[Howard et~al.(2017)Howard, Zhu, Chen, Kalenichenko, Wang, Weyand,
  Andreetto, and Adam]{howard2017mobilenets}
Howard, A.~G., Zhu, M., Chen, B., Kalenichenko, D., Wang, W., Weyand, T.,
  Andreetto, M., and Adam, H.
\newblock Mobilenets: Efficient convolutional neural networks for mobile vision
  applications, 2017.

\bibitem[Ilyas et~al.(2019)Ilyas, Santurkar, Tsipras, Engstrom, Tran, and
  Madry]{ilyas-adv-examples-are-features}
Ilyas, A., Santurkar, S., Tsipras, D., Engstrom, L., Tran, B., and Madry, A.
\newblock Adversarial examples are not bugs, they are features.
\newblock In Wallach, H., Larochelle, H., Beygelzimer, A., d\textquotesingle
  Alch\'{e}-Buc, F., Fox, E., and Garnett, R. (eds.), \emph{Advances in Neural
  Information Processing Systems}, volume~32. Curran Associates, Inc., 2019.
\newblock URL
  \url{https://proceedings.neurips.cc/paper/2019/file/e2c420d928d4bf8ce0ff2ec19b371514-Paper.pdf}.

\bibitem[Microsoft(2022)]{msft-dna}
Microsoft.
\newblock Photodna, 2022.
\newblock URL \url{https://www.microsoft.com/en-us/photodna}.

\bibitem[Roboflow(2021)]{nh-collisions-github}
Roboflow.
\newblock Neuralhash collisions, 2021.
\newblock URL \url{https://github.com/roboflow-ai/neuralhash-collisions}.

\bibitem[Struppek et~al.(2021)Struppek, Hintersdorf, Neider, and
  Kersting]{breaking-nh-struppek}
Struppek, L., Hintersdorf, D., Neider, D., and Kersting, K.
\newblock Learning to break deep perceptual hashing: The use case neuralhash.
\newblock \emph{CoRR}, abs/2111.06628, 2021.
\newblock URL \url{https://arxiv.org/abs/2111.06628}.

\bibitem[Whitley(1994)]{whitley1994genetic}
Whitley, D.
\newblock A genetic algorithm tutorial.
\newblock \emph{Statistics and computing}, 4\penalty0 (2):\penalty0 65--85,
  1994.

\bibitem[Ygvar(2021)]{nh-repro-github}
Ygvar, A.
\newblock Apple neural hash to onnx, 2021.
\newblock URL \url{https://github.com/AsuharietYgvar/AppleNeuralHash2ONNX}.

\end{thebibliography}
